\documentclass[twocolumn,showpacs,preprintnumbers,amsmath,amssymb]{revtex4}

\usepackage{graphicx}
\usepackage{subfigure}
\usepackage{dcolumn}
\usepackage{bm}
\usepackage{upgreek}

\usepackage[colorlinks=true, linkcolor=black, citecolor=black, urlcolor=black]{hyperref}

\newcommand{\bra}[1]{\left\langle#1\right\rvert}
\newcommand{\ket}[1]{\left\lvert #1\right\rangle}

\newcommand{\ME}[3]{\ensuremath{\left \langle \left. #1 \right. \right\lvert \hspace{-0.15em}#2\hspace{-0.15em}\left\rvert \left. #3 \right. \right \rangle}}

\newcommand{\eref}[1]{\mbox{Eq.~(\ref{eqn:#1})}}
\newcommand{\fref}[1]{\mbox{Fig.~\ref{fig:#1}}}
\newcommand{\tref}[1]{\mbox{Table~\ref{tab:#1}}}
\newcommand{\sref}[1]{\mbox{Sec.~\ref{sec:#1}}}

\newcommand{\mt}[1]{\text{#1}}

\newcommand{\vect}[1]{\boldsymbol{#1}}
\newcommand{\mat}[1]{\boldsymbol{#1}}
\newcommand{\operator}[1]{\hat{#1}}

\newcommand{\coloneq}{\mathrel{\mathop:}=}

\newcommand{\half}{\ensuremath{\frac{1}{2}}}

\newcommand{\positivelabel}{+}
\newcommand{\negativelabel}{-}

\newcommand{\Sstate}{\ensuremath{6^2\text{S}_{1/2}}}
\newcommand{\Pstate}{\ensuremath{6^2\text{P}_{3/2}}}

\newcommand{\ThreeJSymbol}[6]{\left(\begin{array}{ccc}#1&#2&#3\\#4&#5&#6\end{array}\right)}
\newcommand{\SixJSymbol}[6]{\left\{\begin{array}{ccc}#1&#2&#3\\#4&#5&#6\end{array}\right\}}

\newcommand{\reduceddipole}{\ensuremath{\langle J\lVert\operator{\mu}\rVert J'\rangle}}

\begin{document}
\title{Stimulated Raman transitions via multiple atomic levels}
\author{James Bateman}\email{jbateman@soton.ac.uk}
\author{Andr\'e Xuereb}
\author{Tim Freegarde}
\affiliation{School of Physics and Astronomy, University of Southampton, Southampton, SO17~1BJ, United Kingdom}
\date{\today}

\begin{abstract}
We consider the stimulated Raman transition between two long-lived states via multiple intermediate states, such as between hyperfine ground states in the alkali-metal atoms.  
We present a concise treatment of the general, multilevel, off-resonant case, and we show how the lightshift emerges naturally in this approach.  We illustrate our results by application to alkali-metal atoms and we make specific reference to cesium.  We comment on some artifacts, due solely to the geometrical overlap of states, which are relevant to existing experiments.
\end{abstract}
\maketitle

\section{Introduction}\label{sec:introduction}
The stimulated Raman transition is an extremely powerful tool for laser manipulation of cold atoms and ions.  By coupling long-lived states via, but never populating, radiative states, experimenters can emulate near-ideal two-level quantum systems with no significant decay \cite{BK4,12,AP30}.
This technique has been used to measure sub-linewidth features \cite{thomas82,ringot01} and to construct atomic interferometers which, by exploiting photon recoil, create spatially separated atomic wave packets which are sensitive to gravity \cite{kasevich91,kasevich92} or fundamental constants \cite{weiss93,weiss94}.
The effective two-level system, which emerges from the Raman problem,
can exhibit behavior such as Rabi flopping \cite{allen_and_eberly,14}, can be used for experiments such as Ramsey interferometry \cite{ramsey50,zanon05}, and can provide the qubits for quantum information processing \cite{QC31,schmidtkaler03,blatt08}.
Sequences of Raman pulses can be used to craft arbitrary superpositions in systems with numerous metastable states \cite{law98} and to prepare such systems in particular states prior to coherent manipulation \cite{boozer07}. Raman processes have also been used to cool atomic samples to far below the photon recoil limit \cite{LC18,davidson94,reichel94,boyer04}.

Throughout the literature, when the Raman transition is discussed, the level structure of the atom is often approximated to three levels---two metastable states and one intermediate (radiative) state.  The Raman problem is solved for this prototypical case and then extended, without proof, to include the multilevel structure of the atom by summing over the various possible routes (see, e.g., Ref. \cite[\S2.1]{kasevich92}).  Here, by including multiple routes from the outset, we confirm that this simple approach is correct, show how an expression for the lightshift emerges naturally from this treatment, and show that the system behaves as a two-level system with an effective coupling strength and an effective detuning.

There is much existing work related to this problem.  The three-level (single intermediate state) off-resonant case has been treated \cite{wu96}, there have been extensions to four levels \cite{deng83}, and the general multilevel problem has been recast into ``serial'' and ``parallel'' cases \cite{kyrola87}.  It has never been shown rigorously, however, that the three-level case can be extended in the way so often assumed.  In the following, we use the semiclassical approach but alternatively one might consider the Jaynes--Cummings model~\cite{shumovsky86,shore93}.

This article is structured as follows.  We first describe, in \sref{three_and_multi_level_systems}, the Raman transition in a three-level system and we show how this can be generalized to include multiple intermediate states; details of the lengthly calculation are confined to the appendix.  We then derive, in \sref{offresonance}, expressions for the behavior of the quantum-mechanical amplitudes in the general, off-resonance case.  In \sref{illustration} we show how these results can be applied to alkali-metal atoms, and we conclude in \sref{conclusion}.

\section{Three-Level Systems}\label{sec:three_and_multi_level_systems}
The simplest system in which a Raman transition may be driven is the three-level ``$\Lambda$'' system, illustrated in \fref{lambda_system}, in which two long-lived ground states are coupled via a radiative upper state which, because the single-photon detuning is sufficiently large, is never significantly populated.
\begin{figure}
	\includegraphics[width=80mm]{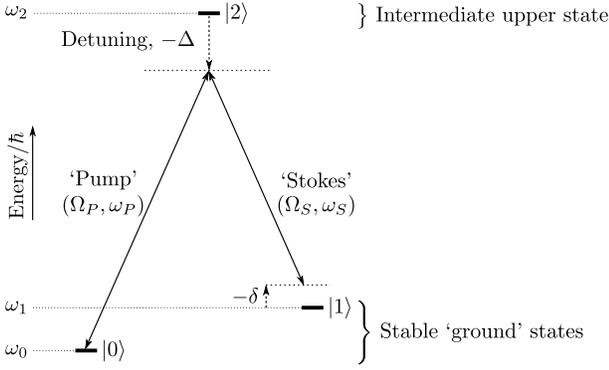}
	\caption{\label{fig:lambda_system}A simple three-level ``$\Lambda$'' system, in which a Raman transition between states $\ket{0}$ and $\ket{1}$ is driven by ``pump'' and ``Stokes'' fields via intermediate state $\ket{2}$. For each field, the coupling strength and frequency are shown in parentheses, and frequencies are chosen to be near two-photon resonance: $\omega_\text{P}=(\omega_2-\omega_0)+\Delta$; $\omega_\text{S}=(\omega_1-\omega_0)+(\Delta+\delta)$.  The single-photon detuning $\Delta$ is large compared with the couplings, $|\Delta|\gg\Omega_\text{P,S}$, and, in this illustration, is negative: $\Delta<0$. The two-photon detuning $\delta$ is small compared with the separation between the ground states and, in this illustration, is also negative.}
\end{figure}
We label the states of the system by $\ket{n}$, with states $\ket{0}$ and $\ket{2}$ coupled by the ``pump'' field of strength $\Omega_\text{P}$ and frequency $\omega_\text{P}$, and states $\ket{1}$ and $\ket{2}$ coupled by the ``Stokes'' field of strength $\Omega_\text{S}$ and frequency $\omega_\text{S}$.  Using the usual correspondence between bra--ket and column vector notation (see, e.g., Ref. \cite[\S{II-C}]{cohentannoudji77}), the Hamiltonian for this system may be represented by the following matrix (see, e.g., Ref. \cite[\S3.2]{14}):
\begin{equation}
\left(\begin{array}{lll}\omega_0&0&\Omega_\text{P}\cos{\omega_\text{P} t}\\0&\omega_1&\Omega_\text{S}\cos{\omega_\text{S} t}\\\Omega_\text{P}\cos{\omega_\text{P} t}&\Omega_\text{S}\cos{\omega_\text{S} t}&\omega_2\end{array}\right)\,\text{.}
\end{equation}Here, as in standard treatments, we assume there is no coupling between states $\ket{0}$ and $\ket{2}$ by the Stokes field, or between states $\ket{1}$ and $\ket{2}$ by the pump field.

This Hamiltonian can be simplified by making the rotating wave approximation and transforming to the interaction picture.  This yields a slowly varying Hamiltonian,
\begin{equation} 
\operator{H} = \left(\begin{array}{ccc}0&0&\half\Omega_\text{P}\\0&0&\half\Omega_\text{S}e^{+i\delta t}\\\half\Omega_\text{P}&\half\Omega_\text{S}e^{-i\delta t}&-\Delta\end{array}\right) \,,
\end{equation}
where the pump frequency $\omega_\text{P} = (\omega_2 - \omega_0) + \Delta$ is detuned from single-photon resonance by $\Delta$, and the difference between the pump frequency and the Stokes frequency $\omega_\text{S} = (\omega_1 - \omega_0) + (\Delta + \delta)$ is offset by $\delta$ from the two-photon resonance $(\omega_1-\omega_0)$.  We now extend this interaction picture to include multiple levels and define the Hamiltonian $\operator{H}_\text{A}$ to be
\newcommand{\ignore}[1]{}
\begin{equation}\label{eqn:multistateinteractionpictureH_A}
\left(\begin{array}{ccccc}
0&	0&	\half\Omega_\text{P;2}\ignore{e^{+i\delta t}}              &	\half\Omega_\text{P;3}\ignore{e^{+i\delta t}}              &	\ldots\\
0&	0&	\half\Omega_\text{S;2}e^{+i\delta t}&	\half\Omega_\text{S;3}e^{+i\delta t}&	\ldots\\
\half\Omega_\text{P;2}^*&		\half\Omega_\text{S;2}^*e^{-i\delta t}&		-\Delta_2&	0&	\ldots\\
\half\Omega_\text{P;3}^*&		\half\Omega_\text{S;3}^*e^{-i\delta t}&		0&	-\Delta_3&	\ldots\\
\vdots&		\vdots&		\vdots&	\vdots&	\ddots
\end{array}\right)\text{.}
\end{equation}
The second part of the subscript, $2,3,\ldots,N$, denotes the level to which the pump or Stokes field couples.  Note that the oscillation frequency is the same for each Stokes term because this depends on the difference in the frequency of the fields and not on the Bohr energy of the intermediate level.  However, the single-photon detunings do depend on the intermediate level Bohr frequencies, but we now make the approximation that the detuning is large compared with the separation of these intermediate levels, and hence $\Delta\coloneq\Delta_2\approx\Delta_3\approx\ldots\approx\Delta_N$.

In this limit, the above Hamiltonian describes a Raman system and we expect to see oscillations of population between the ground states.  We solve the Schr\"odinger equation with this Hamiltonian by using unitary transformations to find a basis where the time evolution of the states is simple and the transformed Hamiltonian is diagonal.
When one makes such a conversion between bases, is it possible to find an equivalent Schr\"odinger equation with a transformed Hamiltonian~\cite{14}: if $\ket{\psi_\text{B}}=\operator{O}_\text{BA}\ket{\psi_\text{A}}$, then $i(\partial/\partial t)\ket{\psi_\text{B}}=\operator{H}_\text{B}\ket{\psi_\text{B}}$ where
\begin{equation}\label{eqn:schrodingertransformation}
 \operator{H}_\text{B} = \operator{O}_\text{BA}\left(\operator{H}_\text{A}\operator{O}_\text{BA}^{-1}-i\frac{\partial}{\partial t}\operator{O}_\text{BA}^{-1}\right)\,.
\end{equation}

For the multistate Hamiltonian $\operator{H}_\text{A}$ in \eref{multistateinteractionpictureH_A}, we choose the operator $\operator{O}_\text{BA}$ to be the matrix of eigenvectors; the first term $\operator{O}\operator{H}\operator{O}^{-1}$ is thus the diagonal matrix of the eigenvalues.  In the appendix, we detail a procedure to find the eigensystem of this Hamiltonian; we find two eigenvectors which are superpositions of these ground states, and $N-2$ more which are superpositions of the remaining intermediate levels.  These $N-2$ upper states are decoupled from the ground states and so we ignore them in the following treatment.

The difference between the eigenvalues for the two ground state eigenvectors is
\begin{equation}
 \widetilde{\Omega}_\text{B}=\frac{1}{2\Delta} \; \sqrt{\lvert\vect{\Omega_\text{P}}\cdot\vect{\Omega_\text{S}}^*\rvert^2 + \frac{1}{4}\left(\lVert\vect{\Omega_\text{S}}\rVert^2 - \lVert\vect{\Omega_\text{P}}\rVert^2\right)^2}\,\text{,}
\end{equation}
where, for conciseness, we have represented the couplings as vectors $\vect{\Omega_\text{P}}$ and $\vect{\Omega_\text{S}}$ with components $\Omega_\text{P;i}$ and $\Omega_\text{S;i}$ respectively, and have used vector notation for dot products and norms. This oscillation frequency is composed of a coupling strength $\Omega_\text{B}$ and a detuning $\Delta_\text{B}$, via $\widetilde{\Omega}_\text{B}=\sqrt{\Omega_\text{B}^2+\Delta_\text{B}^2}$, analogously to a two-level system:
\begin{eqnarray}
 \Omega_\text{B}&=&\frac{\lvert\vect{\Omega_\text{P}}\cdot\vect{\Omega_\text{S}}^*\rvert}{2\Delta}\,,\label{eqn:couplingstrength}\\
 \Delta_\text{B}&=&\frac{\lVert\vect{\Omega_\text{S}}\rVert^2 - \lVert\vect{\Omega_\text{P}}\rVert^2}{4\Delta}\label{eqn:lightshift} \,.
\end{eqnarray}
The detuning $\Delta_\text{B}$ is readily identified as the lightshift and we justify this in the next section.

The operator which describes the transformation from the bare ground states to these dressed ground states can, because of normalization, be written as a rotation:
\begin{equation}\label{eqn:rotation}
\operator{O}_\text{BA}=
\left(\begin{array}{rr}\phantom{+}\cos{\theta}&e^{+i\delta t}\sin{\theta}\\-e^{-i\delta t}\sin{\theta}&\cos{\theta}\end{array}\right)\,\text{,}
\end{equation} and the angle $\theta$ is defined by $\tan{\theta}=\left(\Delta_\text{B}-\widetilde{\Omega}_\text{B}\right)/\Omega_\text{B}$.

This treatment is sufficient for the on two-photon resonance case, where $\delta=0$, but in general the effective Hamiltonian $\operator{H}_\text{B}$ also contains a time-derivative second term, originating from the time-dependence of the operator $\operator{O}_\text{BA}$.  Away from the two-photon resonance, where $\delta\neq0$, we find the following slowly varying, but nevertheless time-dependent, effective Hamiltonian:
\begin{equation}\label{eqn:defH_B}
 \operator{H}_\text{B} = \left(\begin{array}{cc}-\delta\sin^2\theta&-\delta e^{+i\delta t} \cos\theta \sin\theta\\-\delta e^{-i\delta t} \cos\theta \sin\theta&\delta\sin^2\theta + \widetilde{\Omega}_\text{B}\end{array}\right)\,.
\end{equation}

\section{Detuning from resonance}\label{sec:offresonance}
The Hamiltonian $\operator{H}_\text{B}$ in \eref{defH_B} has the same form as that for the simple two-level problem in the interaction picture and with the rotating-wave approximation.  We can, therefore, use familiar tools to solve this problem.  First, we transform to find a time-independent Hamiltonian using the operator $\operator{O}_\text{CB}$ in \eref{schrodingertransformation}:
\begin{multline}
 \operator{O}_\text{CB} = \left(\begin{array}{cc}1&0\\0&e^{i\delta t}\end{array}\right)\\
\implies \operator{H}_\text{C} = \left(\begin{array}{cc}-\delta\sin^2\theta&-\delta \cos\theta \sin\theta\\-\delta \cos\theta \sin\theta&\delta\cos^2\theta + \widetilde{\Omega}_\text{B}\end{array}\right)\,.
\end{multline}
Next, analogously to the dressed-states approach, we rotate by an angle $\theta_2$ (thus defining $\operator{O}_\text{DC}$) where
\begin{equation}
 \tan(2\theta_2) = \frac{\delta\sin(2\theta)}{\widetilde{\Omega}_\text{B}-\delta\cos(2\theta)}\,,
\end{equation}
to find a diagonal Hamiltonian $\operator{H}_\text{D}$.  The difference between the diagonal elements of $\operator{H}_\text{D}$ corresponds to the phase evolution frequency $\widetilde{\Omega}_\text{D}$ of the states in this basis.  As in the previous section, we see that this oscillation frequency is composed of a coupling strength $\Omega_\text{B}$ and a modified effective detuning $\Delta_\text{D}$:
$\widetilde{\Omega}_\text{D} = \sqrt{\Omega_\text{B}^2 + \Delta_\text{D}^2}$,
where $\Delta_\text{D}=\Delta_\text{B}-\delta$ is the detuning relative to $\Delta_\text{B}$, which was previously identified as the lightshift.

We now relate the pure phase evolution in this doubly dressed basis to the evolution of the bare states by concatenating the transformations that led us to this final Hamiltonian:
\begin{equation}
 \operator{O}_\text{DA} = \operator{O}_\text{DC} \cdot \operator{O}_\text{CB} \cdot \operator{O}_\text{BA} 
\end{equation}
and $\ket{\psi_\text{D}}=\operator{O}_\text{DA}\ket{\psi_\text{A}}$, or
\begin{equation}\label{eqn:finaltransformation}
\left(\begin{array}{c}D_0\\D_1\end{array}\right)=
\left(\begin{array}{ll}\phantom{+}\cos(\theta+\theta_2)&e^{i\delta t}\sin(\theta+\theta_2)\\-\sin(\theta+\theta_2)&e^{i\delta t}\cos(\theta+\theta_2)\end{array}\right)
\left(\begin{array}{c}A_0\\A_1\end{array}\right)\,,
\end{equation}
where $D_{0,1}$ and $A_{0,1}$ are the ground- and excited-state components of the doubly dressed wave function $\ket{\psi_\text{D}}$ and the bare (interaction picture) wave function $\ket{\psi_\text{A}}$, respectively.  Finally, the time evolution of the doubly dressed states is simply
\begin{equation}\label{eqn:phase_only}
 \left(\begin{array}{c}D_0(t)\\D_1(t)\end{array}\right)=
 \left(\begin{array}{l}D_0(t=0)\\D_1(t=0)e^{i\widetilde{\Omega}_\text{D}t}\end{array}\right)\,.
\end{equation}

Using \eref{finaltransformation} we can find the dressed state initial conditions $D_{0,1}(t=0)$ in terms of the bare state initial conditions $A_{0,1}(t=0)$.  Using these values, we can then use \eref{phase_only} to find the dressed state coefficients at some later time.  Finally, we can invert the transformation in \eref{finaltransformation} to find the time evolution of the bare state amplitudes.

\subsection{Explicit forms of the amplitudes}
The time dependence of the bare state coefficients $A_{0,1}(t)$ is readily calculable from the procedure described above and is stated here for completeness:
\newcommand{\tmp}{\left(\tfrac{1}{2}\widetilde{\Omega}_\text{D} t\right)}
\newcommand{\tmpD}{\frac{\Delta_\text{D}}{\widetilde{\Omega}_\text{D}}}
\newcommand{\tmpO}{\frac{\Omega_\text{D}}{\widetilde{\Omega}_\text{D}}}
\begin{subequations}\label{eqn:explicitA01t}
\begin{align}
 A_0(t)&=\Biggl(A_0(0)\left[\cos\tmp-i\tmpD\sin\tmp\right]\nonumber\\
&\phantom{=}\quad~~+A_1(0)\, i\tmpO\sin\tmp\Biggr)\,;\phantom{e^{-i\delta t}}\\
 A_1(t)&=\Biggl(A_1(0)\left[\cos\tmp+i\tmpD\sin\tmp\right]\nonumber\\
&\phantom{=}\quad~~+A_0(0)\, i\tmpO\sin\tmp\Biggr)e^{-i\delta t}\,.
\end{align}\end{subequations} Hence the system behaves as a two-level system with the coupling strength $\Omega_\text{D}=\Omega_\text{B}$ and detuning $\Delta_\text{D}=\Delta_\text{B}-\delta$, relative to the effective detuning $\Delta_\text{B}$. This justifies our previous identification of $\Delta_\text{B}$ with the lightshift.

\subsection{Oscillation amplitude}
The complete, but cumbersome, formulas in \eref{explicitA01t} describe the behavior of the bare-state amplitudes in terms of the bare state initial conditions.  If, instead, we express this evolution in terms of the initial values in the doubly dressed basis, we see clearly that the evolution is composed of a time-independent offset and an oscillation:
\begin{equation}
 A_0(t) = \cos(\theta+\theta_2)D_0(0)-\sin(\theta+\theta_2)D_1(0)e^{i\widetilde{\Omega}_\text{D}t}\,.
\end{equation}
The population $p_0(t)=\left|A_0(t)\right|^2$ in state $\ket{0}$ therefore oscillates with peak-to-peak amplitude no greater than $m=\sin\left[2(\theta+\theta_2)\right]$ which, expressed in terms of the effective coupling strength and detuning, is
\begin{equation}
 m=\frac{\Omega_\text{B}}{\sqrt{\Omega_\text{B}^2+\Delta_\text{D}^2}}\,.
\end{equation}  This envelope function describes a power-broadened Lorentzian, centered on the light shifted frequency difference between the ground states.  This expression represents the maximum possible population transfer, and any oscillation will be contained within this envelope.

\subsection{Comments}\label{sec:sub_limitations}
A few specific cases are provided here for illustration.  First, for $\vect{\Omega_\text{P}}=(0,0,\Omega_\text{P})$ and $\vect{\Omega_\text{S}}=(0,0,\Omega_\text{S})$ we recover the well-known results for the three level problem.
Next we note two interesting cases: for $\lVert\vect{\Omega_\text{P}}\rVert=\lVert\vect{\Omega_\text{S}}\rVert$, the lightshift $\Delta_\text{B}$ is zero.
On the other hand, for $\lvert\vect{\Omega_\text{P}}\cdot\vect{\Omega_\text{S}}^*\rvert=0$, the Rabi frequency $\Omega_\text{B}$ is zero.

If, as in this last case, the coupling vectors $\vect{\Omega_\text{P}}$ and $\vect{\Omega_\text{S}}^*$ are orthogonal, then the transition is not driven.
Examples include the trivial case where there is no intermediate state to which both ground states are coupled and the case where there are states to which both are coupled, but where these individual coupling strengths sum to zero.  However, unless the vectors are orthogonal, 
it is possible to adjust the pump and Stokes field strengths to ensure the norms of the vectors are equal, and hence that the lightshift is zero.

Our approach relies on the slow time-dependence of the interaction-picture Hamiltonian: we require that the system is near two-photon resonance, as previously stated, and that there is no coupling of state $\ket{0}$ by the Stokes field or of state $\ket{1}$ by the pump field.  If present, these cross-coupling terms would cause the off-diagonal terms in $\operator{H}_\text{A}$ to oscillate in amplitude as well as phase, and the treatment in the appendix would no longer be valid.
 
For our treatment to be valid, it must therefore be possible to identify clearly which field is resonant with which transition (see Refs \cite[\S13.1]{BK4} and \cite[\S3.9]{14} for further discussion).  First, the detuning must be such that no field is close to a single-photon resonance: $\left|\Delta\right|\gg\lVert\vect{\Omega}_\text{P,S}\rVert$ and $\left|\Delta\pm\omega_{10}\right|\gg\lVert\vect{\Omega}_\text{P,S}\rVert$.  Additionally, the coupling strength must be sufficiently small that each ground state can be resolved: $\omega_{10} \gg \lVert\vect{\Omega}_\text{P,S}\rVert$.
If this last condition is violated, the system may still appear Ramanlike and exhibit coherent behavior, but it is not described adequately by the treatment in this article.

\section{Alkali-metal atoms}\label{sec:illustration}
We are able to calculate the Rabi frequency and lightshift for two states $\ket{0}$ and $\ket{1}$ coupled via a number of upper states.  A common embodiment of this situation is the coupling of two ground hyperfine states via a manifold of upper hyperfine states in an alkali-metal atom. 
Indeed, there typically exists many such pairs of states, but, as ensured by conservation of angular momentum, one state is Raman coupled to at most one other state; hence, the total system may be treated as a collection of independent pairwise couplings.

As a typical example, consider the Raman transition between the ground hyperfine states, via the radiative upper states, in atomic cesium.  The pump and Stokes fields, both tuned near to the $D_2$ transition at $852~\text{nm}$, couple states $\ket{\Sstate,F=3}$ and $\ket{\Sstate,F=4}$, respectively, to the \Pstate~manifold, and have a frequency difference near to the hyperfine splitting of $9.2\,\text{GHz}$~\cite{Rb3}. The two ground states and the intermediate states are
\begin{equation}
\begin{array}{lcl}
\ket{0}&=&\ket{\Sstate;F=3;m_F} \,\text{,}\\
\ket{1}&=&\ket{\Sstate;F=4;m_F+q_\text{P}-q_\text{S}}\text{, and}\\
\ket{n}&=&\ket{\Pstate;F=2,3,4,5;m_F+q_\text{P}} \,\text{,}\\
\end{array}
\end{equation}
where $q_\text{P,S}=0,-1,+1$ are the polarizations of the co-propagating pump and Stokes fields and correspond to linear and left and right circular polarizations, respectively; $m_F$ labels the Zeeman sublevel, corresponding to the projection of the total angular momentum $F$ along the quantization axis provided by an external magnetic field~\cite{BK5}. Linear polarization, in this context, refers specifically to the case of the light electric field parallel to the quantization axis; if these axes are orthogonal, the light field interacts with the atom as though it were a superposition of left and right circular polarizations.

The coupling strengths $\Omega_{\text{P,S}}$ depend not only on the light intensities but also on the dipole matrix element for the transition.  Using the Wigner-Eckart theorem~\cite{brink_satchler,BK5,meunier87} we can split the overlap integral needed to find this dipole matrix element and extract from it a purely geometrical term $G$, leaving a term which embodies the other physical details of the transition:
\begin{multline}
\ME{\Pstate, F', m_F'}{\operator{\mu}}{\Sstate, F, m_F}=\reduceddipole\\
\times G\left(I,J,F,m_F,J',F',m_F',q\right)\,\,\text{.}
\end{multline}
The ``reduced'' matrix element, denoted by double bars $\Vert$, depends on many details of the atom, including nuclear mass, and is not easily calculated; it can, however, be found experimentally from measurements of the upper-state lifetime, as described by Loudon~\cite[Eq. (2.57)]{loudon83} and Demtr\"oder~\cite{demtroder03}:
\begin{equation}
 \Gamma = \frac{16\pi^3}{3\epsilon_0 h\lambda^3}\frac{2J+1}{2J'+1}
\left|\reduceddipole\right|^2\mt{.}
\end{equation}
The second part is the product of geometrical terms:
\begin{multline}\label{eqn:geometricalME}
G=(-1)^{2F'+J+I+m_F}\sqrt{\left(2F'+1\right)\left(2F+1\right)\left(2J+1\right)}\\
\times\ThreeJSymbol{F'}{1}{F}{m_F'}{q}{-m_F}\SixJSymbol{J}{J'}{1}{F'}{F}{I}
\end{multline}
where the arraylike symbol in parentheses $(\ldots)$ is the Wigner 3-$j$ symbol and the similar term in braces $\{\ldots\}$ is the Wigner 6-$j$ symbol~\cite[\S 3.3]{csdata}; both are closely related to the Clebsch-Gordan coefficients.  This relation is described in detail by Edmonds~\cite{edmonds96}.  The two states coupled by the Raman interaction are both in $\Sstate$ so, in the calculations that follow, it is only this geometrical term which is relevant.

We imagine the atom in a region of uniform magnetic field and consider an experiment where it is possible to adjust the frequency difference in order to sweep across transitions between various Zeeman sublevels.  The properties which affect the dipole moment are the various quantum numbers: the nuclear spin $I=7/2$; the electron angular momentum $J=1/2$ or $J'=3/2$; the total angular momentum $F=3$ to $F=4$ via $F'=2,3,4,5$; and the aforementioned projection $m_F$ of $F$ along the quantization axis.

The coupling strength for a dipole transition between states $\ket{n}$ and $\ket{m}$ is proportional to the electric field~\cite{BK4}:
\begin{equation}
\Omega_\text{P,S}=E_\text{P,S}\bra{n}\operator{\mu}\ket{m}/\hbar
\end{equation}
where $\operator{\mu}$ is the dipole operator. As above, we can extract a geometrical term and, using the vector notation,
\begin{equation}
\vect{\Omega}_\text{P,S}=E_\text{P,S}\reduceddipole\vect{G}_\text{P,S}/\hbar\,.
\end{equation}
Hence, the relative properties of each of the Zeeman sublevels are determined by the geometrical terms $\vect{G}_\text{P,S}$.  If we revisit the equations for the coupling strength (\eref{couplingstrength}) and the lightshift (\eref{lightshift}), we see that these terms appear as $\left|\vect{G}_\text{P}\cdot\vect{G}_\text{S}\right|$ and $\lVert\vect{G}_\text{P,S}\rVert^2$, respectively. Thus:
\begin{eqnarray}
 \Omega_\text{B}\!\!&=&\!\!\frac{\reduceddipole^2}{2\Delta\,\hbar^2} \left|E_\text{P} E_\text{S}^*\right| \left|\vect{G}_\text{P}\cdot\vect{G}_\text{S}\right|\,\text{and}\\
 \Delta_\text{B}\!\!&=&\!\!\frac{\reduceddipole^2}{4\Delta\,\hbar^2} \Big(|E_\text{S}|^2\lVert\vect{G}_\text{S}\rVert^2 - |E_\text{P}|^2\lVert\vect{G}_\text{P}\rVert^2\Big).\label{eqn:lightshift_Gs}
\end{eqnarray}

It is simple to calculate these factors for a given initial state and pair of polarizations to examine how the lightshift and the coupling strength depend on the strength of the applied fields. We find that $\lVert\vect{G}\rVert^2$, for transitions from $\ket{J=1/2,F=I\pm1/2,m_F}$ to $\ket{J',F',m_F+q}$, driven by light with polarization $q$, are, for an alkali-metal atom with nuclear spin $I$, given by
\newcommand{\fudge}{A}
\begin{equation}\label{eqn:niceG-squared}
 \lVert\vect{G}\rVert^2=\frac{1}{3}\left(1\pm\fudge(J')\,\frac{q\,m_F}{2I+1}\right)
\end{equation}
where $\fudge(1/2)=-2$ for the $D_1$ transition and $\fudge(3/2)=1$ for the $D_2$ transition. The pump vector $\vect{G}_\text{P}$ couples from the lower hyperfine state and corresponds to the negative branch ($F=I-1/2$); the stokes vector $\vect{G}_\text{S}$ couples from the upper hyperfine state and corresponds to the positive branch ($F=I+1/2$). Using these expressions and \eref{lightshift_Gs}, one may easily calculate the lightshift for a given Raman transition in any alkali-metal atom.

We now turn to the coupling strength $\Omega_\text{B}$ which, for equal polarizations $(q_\text{P},q_\text{S})=(1,1)$, is symmetrical about $m_F=0$.  We find
\begin{equation}\label{eqn:niceGpGs11}
 \left|\vect{G}_\text{P}\cdot\vect{G}_\text{S}\right| = \frac{\left|\fudge(J')\right|}{3(2I+1)} \sqrt{(I+1/2)^2-m_F^2}\,,
\end{equation}
for the $D_1$ and $D_2$ transitions, where $\fudge$ is given above. If we now break this symmetry by choosing, for example, $(q_\text{P},q_\text{S})=(0,1)$, we find
\begin{equation}\label{eqn:niceGpGs01}
 \left|\vect{G}_\text{P}\cdot\vect{G}_\text{S}\right| = \frac{\left|\fudge(J')\right|}{3(2I+1)} \sqrt{T(I+1/2-m_F)}\,,
\end{equation}
where $T(n)$ is the $n^\text{th}$ triangular number ($1,3,6,10\ldots$). The dependence of coupling strength on $m_\text{F}$ level is illustrated in \fref{dependence}.

\begin{figure}
	\includegraphics[angle=-90,width=80mm]{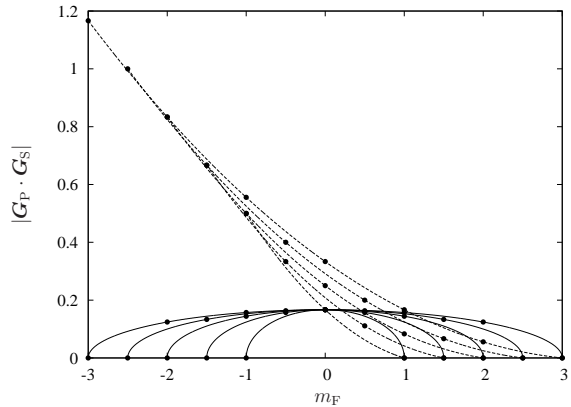}
	\caption{\label{fig:dependence}Illustration of the dependence of coupling strength described by Eqs (\ref{eqn:niceGpGs11}) (solid) and (\ref{eqn:niceGpGs01}) (dashed).  The values have physical meaning at integer and half-integer $m_\text{F}$ only (dots); continuous lines are shown to guide the eye.  A line is shown for each of several nuclear spins, beginning at $I=1/2$ and increasing in steps of one half out from the center (solid) and from bottom to top (dashed). The strongest coupling is between extremal $m_\text{F}$ states by linear and circular polarisations.}
\end{figure}

\newcommand{\fracmod}[1]{#1/24~~}
\newcommand{\fracmods}[1]{\fracmod{\sqrt{#1}}}
\newcommand{\z}{\phantom{+}0}
\begin{table}
\begin{tabular}{c|rrr}
$m_F$&$\left|\vect{G}_\text{P}\cdot\vect{G}_\text{S}\right|$&$\lVert\vect{G}_\text{S}\rVert^2$&$\lVert\vect{G}_\text{P}\rVert^2$\\
\hline
$-3$&$\fracmods{ 7}$&$\fracmod{ 5}$&$\fracmod{11}$\\
$-2$&$\fracmods{12}$&$\fracmod{ 6}$&$\fracmod{10}$\\
$-1$&$\fracmods{15}$&$\fracmod{ 7}$&$\fracmod{ 9}$\\
$\z$&$\fracmods{16}$&$\fracmod{ 8}$&$\fracmod{ 8}$\\
$+1$&$\fracmods{15}$&$\fracmod{ 9}$&$\fracmod{ 7}$\\
$+2$&$\fracmods{12}$&$\fracmod{10}$&$\fracmod{ 6}$\\
$+3$&$\fracmods{ 7}$&$\fracmod{11}$&$\fracmod{ 5}$
\end{tabular}
\caption{Scaling of the coupling strength and the lightshift for the transition $\ket{F=3,m_F}$ to $\ket{F=4,m_F+q_\text{P}-q_\text{S}}$ for $(q_\text{P},q_\text{S})=(1,1)$, in cesium, in terms of the geometrical parts of the dipole matrix elements, as described by Eqs~(\ref{eqn:niceG-squared}) and (\ref{eqn:niceGpGs11}).}\label{tab:F2mf_F3mf_qP=1_qS=1}
\end{table}

The values for a common arrangement are shown in \tref{F2mf_F3mf_qP=1_qS=1}.  While of course linear in any overall scaling of the intensity, the lightshift has a different dependence on the individual field strengths for each of the Zeeman sublevels.  It is offset from zero (for unequal intensities) and is linear in $m_F$; the lightshift between the hyperfine ground states hence has the same dependence on $m_F$ as the Zeeman shift, and, in, e.g., Ref.~\cite{AP33}, is sufficient to account for the majority of the observed spacing between the spectral peaks.

The coupling strengths are not necessarily symmetrical about $m_F=0$ and, for an experiment in which damping is important, this may be manifest as a change in the amplitude of the peaks.  However, the common arrangement of equal polarizations does not show this asymmetry, and, as noted in Ref.~\cite{AP33}, asymmetry in the preparation of the initial states is also important.

\section{Conclusions}\label{sec:conclusion}
We have described the stimulated Raman transition and, by including multiple intermediate states from the outset, have obtained results which give the coupling strength for the multistate system and from which the lightshift naturally emerges. We have applied this method to the cesium atom, given more general expressions for the alkali-metal atoms, and noted how the linear dependence of the lightshift on the $m_F$ level mimics a Zeeman shift. We comment on the possibility of the dependence of the coupling strength on $m_F$ level manifesting as a variation in peak height in experimental spectra.

Our results were derived in the limit of far detuning and our calculations were simplified greatly by this enforced absence of decoherence.  However, for specific coupling strengths and detunings, the problem of finding eigenvalues and vectors can be treated numerically, and many efficient algorithms exist for this task.  Hence, a similar approach might be used for situations including coherent population trapping \cite{gray78,alzetta79} and electromagnetically induced transparency \cite{boller91,fleischhauer05}.


\begin{acknowledgements}
We thank Steven Barnett and Dieter Meschede and his group for helpful discussions.
This work was supported by the UK Engineering and Physical Sciences Research Council (EPSRC) grants EP/E058949/1 and EP/E039839/1, and the European Science Foundation's EuroQUAM project {\em Cavity-Mediated Molecular Cooling}.
\end{acknowledgements}

\appendix*\section{Finding the eigensystem of the multilevel Raman Hamiltonian}\label{derivation}
\newcommand{\dl}{\delta-\lambda}
In this appendix, we find the eigenvalues and eigenvectors for the matrix $\mat{H}$ representing the Hamiltonian $\operator{H}_\text{A}$ of our multilevel system, as described in Sec.~\ref{sec:three_and_multi_level_systems}. 
For brevity in the derivation, we make the replacements $x_n=\half\Omega_{P,n}/\Omega_0$, $y_n=\half\Omega_{S,n}e^{i\delta t}/\Omega_0$, and $\delta=\Delta/\Omega_0$. The frequency $\Omega_0$ is, conceptually, the natural frequency scale for the problem.

\subsection{Determinant}
We calculate the determinant of the $N\times N$ matrix $\mat{A}=\mat{H}-\lambda \mat{I}$,
\begin{equation}
\lvert\mat{A}\rvert=
\left\lvert \begin{array}{ccccc}
-\lambda&	0&		x_2&	x_3&	\ldots\\
       0&	-\lambda&	y_2&	y_3&	\ldots\\
x_2^*&		y_2^*&		\dl&	0&	\ldots\\
x_3^*&		y_3^*&		0&	\dl&	\ldots\\
\vdots&		\vdots&		\vdots&	\vdots&	\ddots
\end{array}\right\rvert\,\text{,}
\end{equation}
to find the characteristic equation and hence the eigenvalues $\lambda$ of $\mat{H}$.  If we define $\mat{A}^{ij}$ to be the matrix $\mat{A}$ with row $i$ and column $j$ removed, and $A_{ij}$ to be the element $(i,j)$ of matrix $\mat{A}$, then using expansion by minors,
\begin{eqnarray}
 \lvert\mat{A}\rvert&=&\sum (-1)^n A_{n0} \lvert\mat{A}^{n0}\rvert\nonumber\\
    &=&-\lambda\lvert\mat{A}^{00}\rvert+\sum_{n\ge2}(-1)^n x_n^*\lvert\mat{A}^{n0}\rvert\;\,\text{.}
\end{eqnarray}
We first evaluate the term $\lvert\mat{A}^{00}\rvert$:
\begin{multline}
 \left\lvert\mat{A}^{00}\right\rvert=
\left\lvert \begin{array}{cccc}
-\lambda&	y_2&	y_3&	\ldots\\
y_2^*&		\dl&	0&	\ldots\\
y_3^*&		0&	\dl&	\ldots\\
\vdots&		\vdots&	\vdots&	\ddots
\end{array}\right\rvert\\
=
-\lambda\left\lvert\left(\mat{A}^{00}\right)^{00}\right\rvert
+\sum_{n\ge2}(-1)^n y_n^*~\left\lvert\left(\mat{A}^{00}\right)^{n-1,0}\right\rvert\;\,\text{.}
\end{multline}
As before, we decompose the determinant in terms of the elements in the first column.  The first minor matrix is diagonal $(\mat{A}^{00})^{00}=(\dl)\mat{I}$, and hence the first term is $-\lambda(\dl)^{N-2}$.  For subsequent terms $\lvert(\mat{A}^{00})^{n-1,0}\rvert$; $n\ge2$, we find $-y_n^*y_n(\dl)^{N-3}$, where the problem of calculating the determinant of each minor matrix  is greatly simplified by swapping columns to ensure each is upper diagonal with diagonal elements $(y_n, \dl, \dl, \ldots)$.  Overall, these terms sum to $-\lVert\vect{y}\rVert^2(\dl)^{N-3}$.  Hence,
\begin{equation}
 \lvert\mat{A}^{00}\rvert=-\lambda(\dl)^{N-2}-\lVert\vect{y}\rVert^2(\dl)^{N-3}\,\text{.}
\end{equation}
We now consider the terms $\lvert\mat{A}^{n0}\rvert$ for $n\ge2$:
\begin{equation}
\mat{A}^{20}= 
\left\lvert \begin{array}{ccccc}
0&		x_2&	x_3&	x_4&	\ldots\\
-\lambda&	y_2&	y_3&	y_4&	\ldots\\
y_3^*&		0&	\dl&	0&	\ldots\\
y_4^*&		0&	0&	\dl&	\ldots\\
\vdots&		\vdots&	\vdots&	\vdots&	\ddots
\end{array}\right\rvert\,\text{.}
\end{equation} Before deconstructing $\mat{A}^{20}$, we note that the next term, $\mat{A}^{30}$, is of the same form if we interchange the columns headed $x_2$ and $x_3$:
\begin{multline}
\mat{A}^{30}= 
\left\lvert \begin{array}{ccccc}
0&		x_2&	x_3&	x_4&	\ldots\\
-\lambda&	y_2&	y_3&	y_4&	\ldots\\
y_2^*&		\dl&	0&	0&	\ldots\\
y_4^*&		0&	0&	\dl&	\ldots\\
\vdots&		\vdots&	\vdots&	\vdots&	\ddots
\end{array}\right\rvert\\
=
-\left\lvert \begin{array}{ccccc}
0&		x_3&	x_2&	x_4&	\ldots\\
-\lambda&	y_3&	y_2&	y_4&	\ldots\\
y_2^*&		0&	\dl&	0&	\ldots\\
y_4^*&		0&	0&	\dl&	\ldots\\
\vdots&		\vdots&	\vdots&	\vdots&	\ddots
\end{array}\right\rvert\,\text{,}
\end{multline}with a sign change.
The first term of this determinant has prefactor zero and so can be ignored.  For the second term, the minor matrix $(\mat{A}^{20})^{10}$ is upper diagonal and hence has determinant $\left\lvert(\mat{A}^{20})^{10}\right\rvert=x_2(\dl)^{N-3}$ and, similarly, $\left\lvert(\mat{A}^{n0})^{10}\right\rvert=(-1)^n x_n(\dl)^{N-3}$ for $n\ge2$.

We now investigate the determinants $\left\lvert(\mat{A}^{20})^{m0}\right\rvert$ for $m\ge2$.  In the expansion of $\left\lvert(\mat{A}^{20})^{20}\right\rvert$, the only terms which have nonzero coefficient are $\left\lvert((\mat{A}^{20})^{20})^{00}\right\rvert=y_3(\dl)^{N-4}$ and $\left\lvert((\mat{A}^{20})^{20})^{10}\right\rvert=x_3(\dl)^{N-4}$.  We can extend this treatment for $n\ge2$ and $m\ge2$.  Finally, we find
\begin{multline}
 \left\lvert\mat{A}\right\rvert=\lambda^2(\dl)^{N-2} + \lambda(\lVert\vect{x}\rVert^2 + \lVert\vect{y}\rVert^2)(\dl)^{N-3}\\
+\frac{1}{2}\sum_{i,j}\left\lvert x_iy_j-x_jy_i\right\rvert^2(\dl)^{N-4}\,\,\text{.}
\end{multline}

\subsection{Eigenvalues}
The equation $\left\lvert\mat{A}\right\rvert=0$ clearly has solution $\lambda=\delta$ with multiplicity $N-4$.  With this factor removed, and using $\lVert\vect{x}\rVert^2\lVert\vect{y}\rVert^2 - \left\lvert\vect{x}\cdot\vect{y}^*\right\rvert^2=\frac{1}{2}\sum_{i,j}\left\lvert x_iy_j-x_jy_i\right\rvert^2$ to phrase this equation in terms of vectors, we obtain 
\begin{multline}
 \lambda^2(\dl)^{2} + \lambda(\lVert\vect{x}\rVert^2 + \lVert\vect{y}\rVert^2)(\dl)\\
+\lVert\vect{x}\rVert^2\lVert\vect{y}\rVert^2-\left\lvert \vect{x}\cdot\vect{y}^*\right\rvert^2=0\,\text{,}
\end{multline}
which is a fourth-order polynomial in $\lambda$ (with leading coefficient $1$) and hence the product of the remaining roots $\lambda_i$ equals the constant term.  This term is finite, and so \emph{at least} one $\lambda_i$ becomes negligible as $\delta\to\pm\infty$, and hence we can make the approximation $(\dl)\to\delta$ in this limit.  The resulting equation is a quadratic in $\lambda$ with solutions
\begin{equation}
 \lambda_\pm = \frac{-\left(\lVert\vect{x}\rVert^2 + \lVert\vect{y}\rVert^2\right)
\pm\sqrt{\left(\lVert\vect{x}\rVert^2 - \lVert\vect{y}\rVert^2\right)^2 + 4 \lvert\vect{x}\cdot\vect{y}^*\rvert^2}
}{2\delta} \,\text{.}
\end{equation}

We now seek the eigenvectors associated with these two finite eigenvalues.

\subsection{Eigenvectors}
The eigenvalue equation $\mat{H}\vect{a}=\lambda\vect{a}$ yields the following:
\begin{eqnarray}
 \sum_{i\ge2} x_i a_i&=&\lambda a_0\text{;}\label{eqn:eigenvector1}\\
 \sum_{i\ge2} y_i a_i&=&\lambda a_1\text{;}\label{eqn:eigenvector2}\\
 x_i^*a_0 + y_i^*a_1&=&(\lambda-\delta)a_i\,\text{for }i\ge2\,\text{.}\label{eqn:eigenvector3}
\end{eqnarray}
If we multiply \eref{eigenvector3} by $a_i^*$ and sum over $i\ge2$, and then enforce the normalization condition $\sum a_i^*a_i=1$, we arrive at
\begin{equation}
 (\lambda-\delta)\left[1-\left(|a_0|^2 + |a_1|^2\right)\right]=\lambda\left(|a_0|^2 + |a_1|^2\right)\,\text{,}
\end{equation}
and hence
\begin{equation}
 |a_0|^2 + |a_1|^2 = 1-\lambda/\delta\,\text{,}
\end{equation}
which, in the limit of large $\delta$, tends to unity.  By way of confirmation, we see from \eref{eigenvector3} that
\begin{equation}
 a_i=\frac{x_i^*a_0+y_i^*a_1}{\lambda-\delta} \,\text{for } i\ge2\,\text{,}
\end{equation}
which tend to zero in this limit.
The two eigenstates associated with the two finite eigenvalues are hence orthogonal superpositions of the two ground eigenstates; we represent this transformation as a rotation and proceed to find its angle.
Using Eqs~(\ref{eqn:eigenvector1}) and~(\ref{eqn:eigenvector3}), we have
\begin{eqnarray}
 a_0&=&\frac{1}{\lambda}\sum_{i\ge2} x_i a_i\nonumber\\
    &=&\frac{1}{\lambda(\lambda-\delta)}\sum_{i\ge2} x_i \left[x_i^*a_0 + y_i^*a_1\right]\nonumber\\
    &=&\frac{1}{\lambda(\lambda-\delta)}\left[a_0\vect{x}\cdot\vect{x}^* + a_1 \vect{x}\cdot\vect{y}^*\right]\,\text{,}
\end{eqnarray}
and similarly for $a_1$.  Hence we obtain
\begin{equation}
 \frac{a_1}{a_0}=\frac{a_0 \vect{y}\cdot\vect{x}^* + a_1 \vect{y}\cdot\vect{y}^*}{a_0 \vect{x}\cdot\vect{x}^* + a_1 \vect{x}\cdot\vect{y}^*}\,\text{,}
\end{equation}
which, because $\left\lvert a_0\right\rvert^2+\left\lvert a_1\right\rvert^2=1$, 
we can express as the tangent of an angle:
\begin{equation}
 e^{i\phi}\tan{\theta}=a_1/a_0
=\frac{
-\left(\lVert\vect{x}\rVert^2 - \lVert\vect{y}\rVert^2\right)
\pm\chi}{2\vect{x}\cdot\vect{y}^*}\,\text{,}
\end{equation}
where $\chi=\sqrt{4 \lvert\vect{x}\cdot\vect{y}^*\rvert^2 + \left(\lVert\vect{y}\rVert^2 - \lVert\vect{x}\rVert^2\right)^2}$ and $\phi=-\arg(\vect{x}\cdot\vect{y}^*)$.

The transformation from the bare-state basis to this dressed-state basis can hence be described by the rotation
\begin{equation}
\left(\begin{array}{c}\ket{\positivelabel}\\\ket{\negativelabel}\end{array}\right)
=
\left(\begin{array}{rr}\cos{\theta}&e^{+i\phi}\sin{\theta}\\-e^{-i\phi}\sin{\theta}&\cos{\theta}\end{array}\right)
\left(\begin{array}{c}\ket{0}\\\ket{1}\end{array}\right)\,\text{,}
\end{equation}
and oscillations are thus driven with amplitude $m=2\cos{\theta}\sin{\theta}$ at the rate 
\begin{equation}
\widetilde{\Omega}_\text{B}=\Omega_0\left(\lambda_+-\lambda_-\right)=\frac{\Omega_0^2}{\Delta}\chi \,\text{.}
\end{equation}

We identify the effective coupling strength $\Omega_\text{B}$ and detuning $\Delta_\text{B}$ in terms of the angle $\theta$ defined above:
\begin{eqnarray}
 \Omega_\text{B}&=&\sin{2\theta}~\widetilde{\Omega}_\text{B}\,\text{ and}\\
 \Delta_\text{B}&=&\cos{2\theta}~\widetilde{\Omega}_\text{B}\,\text{.}
\end{eqnarray}

Using the trigonometric identity $\tan{\theta}=\frac{1-\cos{2\theta}}{\sin{2\theta}}$ we obtain
\begin{eqnarray}
 \sin{2\theta}&=&2\left|\vect{x}\cdot\vect{y}^*\right|/\chi\,\text{ and}\\
 \cos{2\theta}&=&\left(\lVert\vect{y}\rVert^2 - \lVert\vect{x}\rVert^2\right)/\chi\,\text{,}
\end{eqnarray}
and, finally,
\begin{eqnarray}
 \Omega_\text{B}&=&\frac{\Omega_0^2}{\Delta} \, 2 \left|\vect{x}\cdot\vect{y}^*\right|\,\text{ and}\\
 \Delta_\text{B}&=&\frac{\Omega_0^2}{\Delta} \,\left(\lVert\vect{y}\rVert^2 - \lVert\vect{x}\rVert^2\right)\text{.}
\end{eqnarray}


\end{document}